# Behavioral Compatibility of Simulink Models for Product Line Maintenance and Evolution

**Bernhard Rumpe**
Software Engineering
RWTH Aachen University,
Germany
http://www.se-rwth.de/

**Christoph Schulze**
Software Engineering
RWTH Aachen University,
Germany
http://www.se-rwth.de/

**Michael von Wenckstern**
Software Engineering
RWTH Aachen University,
Germany
http://www.se-rwth.de/

**Jan Oliver Ringert**
School of Computer Science
Tel Aviv University, Israel
http://www.cs.tau.ac.il/

**Peter Manhart**
Software-Variantenmanagement
Daimler AG, Germany
http://www.daimler.com

## ABSTRACT

Embedded software systems, e.g. automotive, robotic or automation systems are highly configurable and consist of many software components being available in different variants and versions. To identify the degree of reusability between these different occurrences of a component, it is necessary to determine the functional backward and forward compatibility between them. Based on this information it is possible to identify in which system context a component can be replaced safely by another version, e.g. exchanging an older component, or variant, e.g. introducing new features, to achieve the same functionality.

This paper presents a model checking approach to determine behavioral compatibility of Simulink models, obtained from different component variants or during evolution. A prototype for automated compatibility checking demonstrates its feasibility. In addition implemented optimizations make the analysis more efficient, when the compared variants or versions are structurally similar.

A case study on a driver assistance system provided by Daimler AG shows the effectiveness of the approach to automatically compare Simulink components.

## 1. INTRODUCTION

The establishment of a software product line introduces an additional dimension of complexity to the general difficult tasks to design and maintain larger software systems. As consequence maintaining a larger software product line, thereby handling different versions and variants of software components, is an even more complex task. During the evolution and maintenance of a software product line the knowledge about compatibility relations between different versions and variants of specific software components can be very helpful to answer questions like: can a newer version or variant be introduced in an old context? Under which circumstances is it possible to replace a newer but erroneous component by an older version? Does an older component still need to be maintained or can it be replaced in every available system context?

In the context of the automotive domain previous work [29] describes the compatibility information of components (backward, upward, and full compatibility) and their notation in a tabular format. Engineers use this information to derive whether for a specific system configuration compatibility issues regarding interface or functionality arise.

Given two components $C_0$ and $C_1$ with compatible interfaces, $C_1$ is backward compatible to $C_0$ if and only if the behavior of $C_0$ is a subset of $C_1$, i.e., the functionality of $C_0$ is contained in $C_1$. Upward compatibility is the inverse relation of backward compatibility: $C_0$ is upward compatible to $C_1$ if and only if the behavior of $C_0$ is a superset of $C_1$. Full compatibility is provided if both is the case. Since upward compatibility is the inverse of backward compatibility, in the latter we focus only on calculating backward compatibility.

Nevertheless the provided compatibility relations between components had been derived by experts. The manual analysis of behavior compatibility can be error-prone and time-intensive. This paper proposes a model checking approach to pairwise compute compatibility for different versions or variants of components. Interface compatibility is a precondition for behavior compatibility. Engineers can provide a mapping between corresponding compatible ports, and compatibility of the mapping and component interfaces can be checked syntactically. If the two compared components are not backward compatible, then the approach provides input sequences producing different component output sequences. This makes the compatibility answer comprehensible to engineers and helps to identify the source of incompatibility.

The approach is implemented in a prototype that takes as input two Simulink models that may consist of the following Simulink library categories [24]: Discrete, Logic and Bit Operations, and Math Operations as well as Ports & Subsystems and Signal Routing with Boolean, enumeration, and integer variables (may be bundled by bus objects). The prototype produces a compatibility statement and, in case the components are incompatible, an input sequence witnessing the incompatibility. An extension of the tool al-lows to check backward compatibility when the interfaces are not equivalent but the new component has additional inputs that can be replaced by constant values. Three effective optimizations help to efficiently analyze compatibility. These are detection and pruning of common subsystems, separate analyses per outport, and pre-computation of reachable state space.

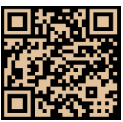

The outline of the paper is as follows: Sect. 2 repeats background of SMT solvers, simulation preorder relations, and state space explosion. It gives an overview of the technologies underlying the prototype implementation: MontiCore [16], MontiArc [15], and MontiArcAutomaton [31]. While Sect. 3 presents a basic approach checking backward compatibility; Sect. 4 describes additionally implemented optimizations and extensions making the basic algorithm more suitable in a product line environment. Sect. 5 evaluates and discusses the concept by testing behavioral compatibility for different components of a driver assistance system provided by Daimler AG. Sect. 6 compares the presented concept to existing approaches used in model-checking tools and for model comparison. The paper finishes with a conclusion of the presented concept and tool chain.

## 2. FOUNDATIONS

This section gives a basic introduction to input/output extended finite automata (I/O-EFAs), the state space explosion problem, the simulation preorder relation, and Satisfiability Modulo Theory (SMT) checking as well as to the tools and frameworks underlying for the developed prototype.

### 2.1 Input/Output Extended Finite Automata

Following [37, 32], an input/output extended finite automaton (I/O-EFA) is a tuple $A = (S, s_0, D, d_0, U, Y, E)$, where

- $S$ is the set of states with initial state $s_0 \in S$,
- $D$ is a set of internal variables with initial values $d_0$,
- $U$ and $Y$ are sets of input and output variables, and
- $E$ is a set of transitions $o_e \xrightarrow[\{y=h_e(d,u); d=f_e(d,u)\}]{[g_e(d,u)]} t_e$ from $o_e \in S$ to $t_e \in S$, with guard condition $g_e \subseteq D \times U$, output function $h_e : D \times U \to Y$, and data update function $f_e : D \times U \to D$ all depending on the current variable values $d \in D$ and input $u \in U$.

I/O-EFAs are a formalism expressive enough for representing Simulink models [37]. Internal variables are introduced for `DataMemory` or `UnitDelay` blocks and also indirectly for predefined components such as `EdgeRising` or `FlipFlop`. The I/O-EFA in [37] also contains symbolic in- and outputs for continuous calculations, e.g. integration with dynamic intervals. Since this paper only handles discretized models, e.g. integration with predefined intervals, the I/O-EFA definition in this paper omits all symbolic variables.

An I/O-EFA without internal variables $D$ and having output functions $h_S$ for each source state instead of every transition is an input/output transition system (I/O-TS) [37].

Transforming an I/O-EFA to an I/O-TS means that every state will be replaced by a set of new states representing every combination of this state with its variable values. The unfolding of variables into the state space (also known as state space explosion problem [4]) is a limiting factor for the type of variables supported by this approach. All variable domains have to be finite.

Without loss of generality, each guard condition in every I/O-EFA and every I/O-TS in this paper will become `true` for at least one input value $u \in U$.

### 2.2 Simulation Preorder Relation

Intuitively, the existence of a simulation preorder between two I/O-TS $A$ and $B$ means that for the same input: (1) the set of all sequences of $B$'s transition executions is a superset of all sequences of $A$'s transition executions, and (2) both having an identical output [12].

Formally, a simulation relation for I/O-TSs $A$ and $B$ with sets of states $S_A$ and $S_B$, and sets of transitions $E_A$ and $E_B$ is a binary relation $R \subseteq S_A \times S_B$, where if $(a, b) \in R$, and transition $a \xrightarrow{[g_a(u)]} a' \in E_A$ is enabled for an input $u \in U$ then $A$ and $B$ produce equivalent output for states $a$ and $b$ ($y_a \equiv y_b$) and there exists a transition $b \xrightarrow{[g_b(u)]} b' \in E_B$ enabled by $u$ such that $(a', b') \in R$. For two deterministic I/O-TS the simulation relation can be computed starting from the initial states.

In this paper a component `C1` is backward compatible to `C0` if for their corresponding I/O-TS $C_1$ and $C_0$ the start states of $C_0$ are in a simulation relation with the start states of $C_1$. This means that `C1` does produce the same output sequence as `C0` for any given input sequence.

Since all Simulink numeric values will be internally stored as a bitsequence, every Simulink simulation execution of discretized models belongs to Mayr's (1, 1)-PRS class [25] of deterministic finite-state process rewrite systems. Therefore (strong) bisimilarity and (deterministic) simulation preorder of two Simulink components are always decidable [1]. To check enabledness of transitions for inputs an algorithm needs to compare guard functions. A convenient means for comparison are SMT solvers.

### 2.3 Satisfiability Modulo Theory solvers

Satisfiability Modulo Theory (SMT) solvers extend propositional satisfiability solvers with specialized theory solvers [27]. Theories relevant for computing simulation relations of I/O-TS of Simulink components are, e.g., (non-) linear arithmetic and integer arithmetic, which allow SMT solvers to check equality of discrete or real-valued functions.

This prototype tool presented here uses Microsoft's Z3 SMT solver, which won the SMT-Comp in 2011[9]. [8] shows that in the category "non-linear integer arithmetic with uninterpreted sort and function symbols as well as quantifiers" (UFNIA) the Z3 solver is the best and fastest one in 2014. The presented prototype uses uninterpreted functions to find variable assignments making SMT expressions valid.

### 2.4 MontiCore, MontiArc, and MontiArcAutomaton

MontiCore is a framework for developing textual domain specific languages [16]. Features of MontiCore include language inheritance and embedding. This allows modularizing and easily reusing already defined languages such as members of the UML/P [33] family.

MontiArc is a domain specific language to model component and connector architectures [15]. In MontiArc a component contains input and output ports as well as a behavior description. Common data types such as `String`, `Integer`, `Boolean`, etc. are supported as data types of ports.

MontiArcAutomaton (MAA) is an extension of MontiArc which allows modeling the component behavior with au-tomata [31]. The MAA syntax allows defining states, variables, and transitions with guards, output, and variable assignments using an expression language similar to Java.

The MAA modeling language follows the I/O$^\omega$ [32] paradigm. The syntax of the MAA language family subsumes the I/O-EFA's as well as the I/O-TS's syntax. Therefore the prototype uses MAA to represent these automata.

## 3. BACKWARD COMPATIBILITY CHECK

Based on the foundations, this section can now describe the basic approach for checking backward compatibility of Simulink components. The approach is implemented in a prototype tool with additional optimizations and advanced features presented in Sect. 4.

The prototype hides the theoretical background on model checking from engineers and developers. Its focus lies on friendly usage. A developer only selects two Simulink models to compare and the prototype carries out the check fully automatically.

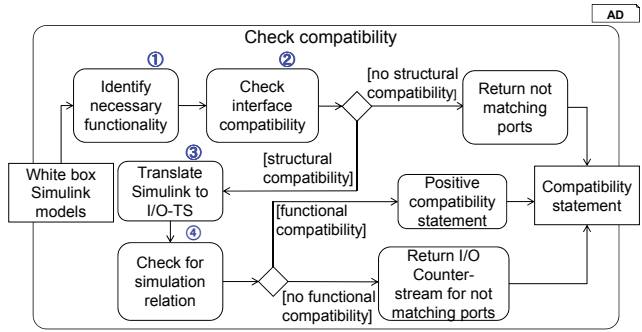

Figure 1: Activity to check backward compatibility.

Fig. 1 gives an overview how backward compatibility statements of Simulink models are derived. First, it is necessary to define the context of the component to identify the required functionality. Based on this information necessary in- and outgoing ports are identified. Second, interface compatibility between the two models is checked: interface compatibility is a necessary condition for behavioral backward compatibility. The prototype assumes the existence of a mapping between ports of the two models. Mappings are mostly computed based on port names.

Third, the prototype translates each Simulink model under comparison to an I/O automaton via a construction inspired by [37] (see Subsect. 3.1 and Subsect. 3.2).

Finally, a detailed behavioral compatibility check is performed (see Subsect. 3.3). The prototype checks for a simulation relation between the I/O automata derived in the previous step. If such a relation does not exist, a counter example for backward compatibility is produced.

### 3.1 Transforming Simulink Models to Control Flow Graphs

The computation of a control flow graph (CFG) is an intermediate step from Simulink models to I/O-EFA automata in the translation of [37]. This step is repeated here to make the paper self-contained and explain the implemented optimizations. The CFG obtained from a Simulink model represents the execution of atomic blocks and the resulting changes of signals.

Simulink's simulation cycle consists of three steps: initialization, output calculation and variable update. Initializa-tion is performed only once, while output calculation and update of internal variables are performed in each simulation cycle for all Simulink blocks with the proper frequency. The transformation identifies atomic Simulink blocks inside the hierarchy and transforms these into a set of calculation steps, one step for each atomic block. To combine different atomic blocks a connection rule is applied. In this step the different states for each single block are concatenated based on Simulink's *sorted order index* [24] of the model. Thereby the simulation cycle is considered by first concatenating the output calculation transitions before the update transitions. To extract enabled subsystems a conditioned rule is applied. This rule introduces more complex branches into the control flow as it introduces nodes whose transitions lead to different target nodes. One cyclic execution of the CFG represents one simulation step of the transformed Simulink model [37].

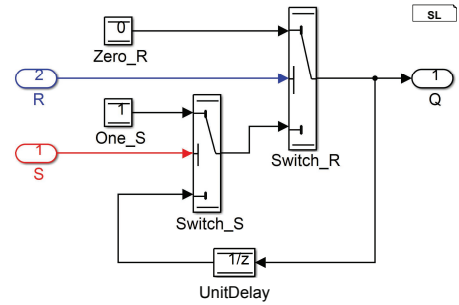

Figure 2: Internal of a `FlipFlop` block (outgoing port `Not_Q` is omitted).

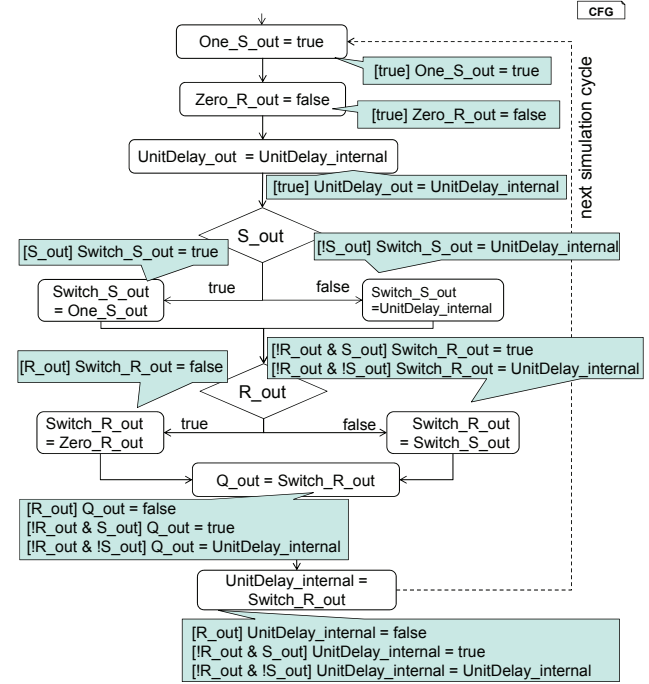

Figure 3: Extracted control flow of Fig. 2's `FlipFlop` based on Simulink simulation scheduling order.

As an example consider the component `FlipFlop` shown in Fig. 2 and its CFG shown in Fig. 3. Nodes in the CFG are labeled with signals they assign. The Simulink scheduler starts with blocks `Zero_R` and `One_S` (upper part of Fig. 2). The CFG starts with assignments to the output signals of these blocks (suffix `_out` in Fig. 3). Next the block `Unit-

`Delay` is executed. Afterwards the CFG branches on the conditional value assignments of block `Switch_S` and afterwards of the block `Switch_R`. Finally, an update of the internal variable `UnitDelay_internal` of block `UnitDelay` is performed in the last node of the simulation cycle.

## 3.2 Transforming CFGs to I/O-Automata

After the CFG extraction, a dependency analysis of assignments in the CFG is used to obtain output and variable assignments at the end of every execution cycle. From these assignments an I/O-EFA is constructed. The I/O-EFA is deterministic and has only a single state. Every execution of a transition corresponds to the execution of a complete cycle in the CFG. The state space of the automaton is encoded via internal variables.

For the CFG of component `FlipFlop` all propagated variable assignments are shown in the comment boxes for each node in Fig. 3. Assignments that depend on inputs or the value of local variables are guarded, e.g., the assignment `Switch_S_out = true` is guarded with the input signal `S_out`. All transitions as well as the outport and internal variable assignments are derived from the bottom most propagated variable assignments (bottom most comment boxes in Fig. 3). In this example, the local variable `UnitDelay_internal` always stores the last output value, and the condition as well as the assignment do not change. In general, conditions and assignments can change or more than one internal variable can be involved calculating the outgoing port. The corresponding state space is encapsulated in the corresponding guard conditions and related internal variable assignments.

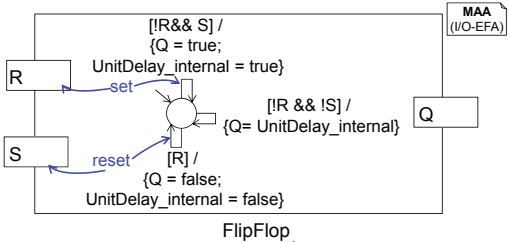

Figure 4: Extracted I/O-EFA of Fig. 3.

The prototype requires an I/O-TS to calculate a simulation relation in the next step of checking backward compatibility. The translation from I/O-EFA to I/O-TS unfolds internal variables into states. In addition state-based output functions are defined based on the outputs in the I/O-EFA.

Since Microsoft's Z3-Solver is able to handle `ite` (if-then-else) expressions, the output function of the resulting I/O-TS is a concatenation of `ite`s where the I/O-EFA's guard condition is in the *if* part, the I/O-EFA's transition expression in the *then* part and the *else* part contains the next `ite`-condition.

## 3.3 Simulation

The algorithm determining the backward compatibility is based on the simulation preorder definition of I/O-TS in Subsect. 2.2. Fig. 5 illustrates this algorithm (`B` is simulated by `A`) consisting of four key parts: ① Testing whether the state pair $(a, b)$ have already been visited by the algorithm; if it is so nothing has to be done. This way the algorithm also terminates for I/O-TS containing loops. ② Checking whether the output function of the states being in relation are equivalent for inputs of `B`. ③ Taking a specific transition of the actual state in `B`, determining possible input values satisfying the guard condition of this transition, and looking for transitions of actual state in `A` being activated by these input values. ④ Checking whether all successor states based on the selected transitions are also in simulation preorder relation; the algorithm calls itself recursively.

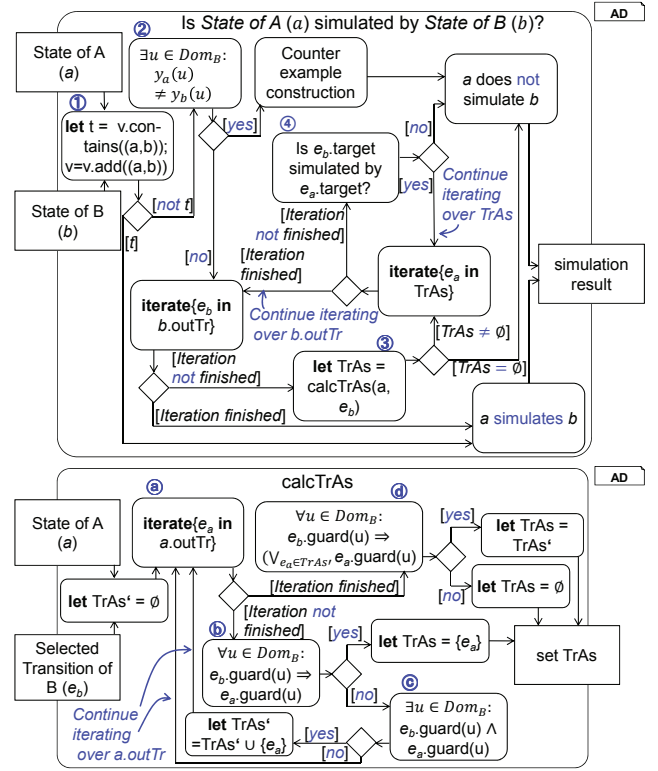

Figure 5: Activity Diagram of Simulation preorder algorithm. Bold words are OCL/P [33] keywords.

In ③ the algorithm finds all outgoing transitions, the set $TrAs$, of the actual state `a` satisfying the following properties: (i) $\forall e_a \in TrAs : e_a \in a.outTr$,

(ii) $e_b.guard \Rightarrow \left( \bigvee_{e_a \in TrAs} e_a.guard \right)$,

(iii) $\forall e'_a \in TrAs : e_b.guard \not\Rightarrow \left( \bigvee_{e_a \in TrAs \setminus \{e'_a\}} e_a.guard \right)$

For deterministic I/O-TSs the set $TrAs$ is unique defined by (i)-(iii), and can be calculated as shown in the bottom activity diagram in Fig. 5.

(i) forces the transitions starting from state $a$ (is ⓐ). (ii) says if the transition $e_b$ is activated by an input $u$, then there must be at least one transition $e_a$ also being activated by $u$ (is ⓓ). (iii) forces that $TrAs$ in (ii) contains only a minimum number of $a$'s outgoing transitions ((iii) is equivalent to condition ⓒ). The optimization ⓑ checks whether there exists a transition $e_a$ covering all activations of $e_b$ directly.

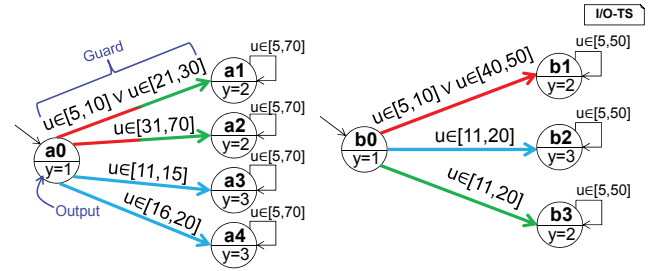

Figure 6: Example explaining part ③ of algorithm.

In Fig. 6 I/O-TS $A$ simulates I/O-TS $B$, and $TrAs(b0 \to b1) = \{a0 \to a1, a0 \to a2\}$, $TrAs(b0 \to b2) = \{a0 \to a3, a0 \to a4\}$, $TrAs(b0 \to b3) = \{a0 \to a1, a0 \to a2\}$. $B$ does *not* simulate $A$, because for the transition $a0 \to a2$ no activiated outgoing transition of $b0$ for the input $u = 70$ can be found (meaning $TrBs(a0 \to a2) = \emptyset$).

Component B (which behavior is represented by I/O-TS B) is simulated by component A if and only if the start state of B ($b_0$) is simulated by the start state of A ($a_0$).

Since the algorithm checks whether $b_0$ is simulated by $a_0$ meaning component B can be replaced by component A, only the input value ranges of B ($u \in Dom_B$) arrive at components A and B. $y_a(u)$ is the output function of a state a under the input values $u \in Dom_B$ ($y_a$ as well as $u$ are vectors).

The counterexample construction creates one possible input stream vector, so that the output streams of both components differ at exactly one time slot. Tracking differences in the Simulink model can be done by using a `SignalBuilder` [24] generating this input stream.

## 4. IMPLEMENTED OPTIMIZATIONS AND EXTENSIONS

The previous section introduced the basic concept how behavioral compatibility between two Simulink components can be evaluated.

This section contains improvements of the basic algorithm implemented in the prototype for checking the compatibility of similar components: either by further development or variation of models. Typically, a common base of the two compared models can be expected. The common part is identified and removed in order to decrease the model's complexity.

Another optimization uses the relation between in- and outgoing ports in the CFG to remove unnecessary calculations and to divide the I/O-EFA calculating all output ports into several subautomatons handling a subset of the outputs.

The section's last part introduces an extension to compare different variants or versions of models where one model's interface is only a subset of the others interface. The approach fixes the values on input ports that are only available in one of the two models (but influence the result of a required output).

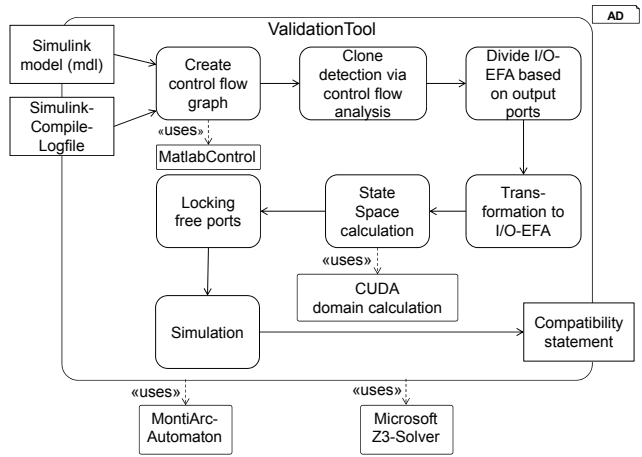

**Figure 7: Implementation: activities to check functional compatibility.**

Fig. 7 shows the complete technical view of the different steps in the currently implemented prototype. All details of the extra steps such as *clone detection*, *divide I/O-EFA based on output ports*, *state space calculation* and *locking free ports* will be explained in the next subsections.

### 4.1 Clone Detection and Pruning

The CFG extracted from a Simulink model is the foundation for a first optimization, which aims to reduce the complexity of the resulting automaton via syntactical or semantical clone detection.

In a first step in the tradition of program slicing [36] the CFG is analyzed to receive the information which CFG parts depend on each other. Therefore the control flow is stepwise traversed and a substitution is performed. This is done by replacing all known parameter by the corresponding equation already traversed, similar to symbolic execution [28]. In addition the related conditions are preserved by performing a logical concatenation with the guard of the actual traversed transition. The greenish-blue comments in squared bubble boxes in Fig. 3 show the substitutions for each variable during the CFG's traversal process of the example from Fig. 2. The substitution of `Switch_R_out` exemplary shows that the complexity of the substitution in general grows exponentially in the branching levels (cf. path explosion [7]).

After the substitution has been performed, the following information can be acquired: (1) Internal variables and ingoing ports which are used to update an internal variable. (2) Internal variables and outgoing ports which are used to update an outgoing port.

In addition all global variables of the Simulink model which are just used to ease the modeling of communication between blocks, and are therefore not representing an internal state of a block are replaced during the substitution. By storing all substitutions in each step, it is possible to compare these substitutions with others extracted from a second model and therefore perform a syntactical clone detection by simply comparing all guard and assignment pairs of one internal place holder, internal variable or outgoing port. There is a high potential for semantical clone detection on this level, which will be discussed in Sect. 5, as it is not part of the current implementation.

If a clone is detected, the clone is removed by introducing a new ingoing port, representing the clone. This ensures that the value will always be the same for both components. Its actual calculation is of no importance for checking compatibility. If the clone is not an internal place holder or internal variable, but an outgoing port, the whole port can be ignored in further steps, as its equivalence and therefore full compatibility has been proven. If, after a clone has been removed, some inputs are not referenced anymore, they are removed. Considering the example from Fig. 8, which represents two variants of one component, a corresponding syntactical clone detection could easily identify the substitution of the outgoing port `VMax_kmh` as a clone. Therefore only the red marked signal flow has to be evaluated any further.

### 4.2 Divide I/O-EFA Based On Output Ports

#### 4.2.1 Theoretical Preliminaries

The inspected Simulink models are completely deterministic. This results in deterministic CFGs and automata.

A trace of an I/O-TS is a sequence of observed input-

145

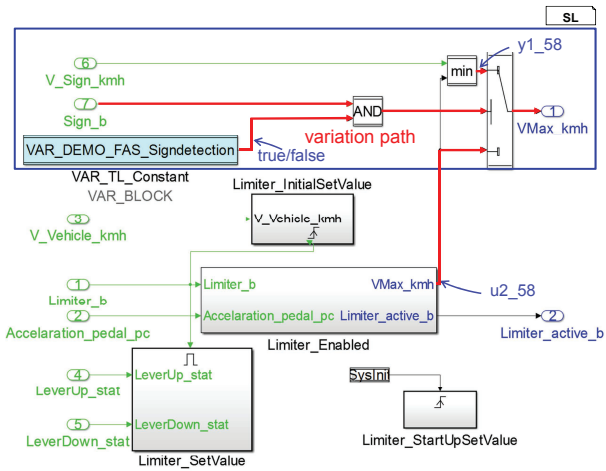

**Figure 8: Identifying syntactical clones based on the variation path.**

output value vectors. If two I/O-TS are not distinguishable by *all* traces they produce, they are trace equivalent.

Trace equivalence is black box equivalence [12] that does not depend on internal states or values of internal variables. Thus, one large automaton calculating several outputs can be divided in multiple automata each calculating one output. This allows a decomposition of the check for trace equivalence of two automata: *iff* all small automata of both components are trace equivalent then the two large automata of both components produce the same result for identical inputs on each outport; thus they are also trace equivalent.

To determine whether two small automata A and B are trace equivalent the algorithm described in Subsect. 3.3 can be used, because simulation equivalence coincides [12] with trace equivalence for deterministic automata.

### 4.2.2 Algorithm

Based on the CFG and evaluating the dependencies between in- and outgoing ports in Subsect. 4.1, it is possible to create for each outgoing port one automaton based on the calculated substitution by building all possible assignment pairs of the outgoing port and its influencing internal variables.

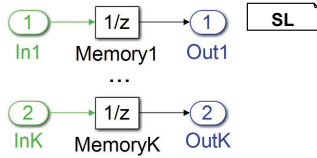

**Figure 9: Theoretical best case dividing automaton.**

In the best case illustrated in Fig. 9, where all of the $k$ output ports are pairwise independent from the other's internal variables and input ports, an automaton division exponentially reduces the state space of the I/O-TS in $\mathcal{O}(2^k)$.

In the worst case where all $k$ output ports depend on *all* internal variables and input ports, the small automatons calculating one output port are exactly the same. Thus all followed operations will be executed $k$ times, and so a linear computation time overhead ($\mathcal{O}(k)$) arises.

The prototype implements these heuristics. More advanced strategies could, e.g. analyze the dependencies between variables, input ports and the outgoing ports after the CFG has been extracted and the substitution is performed.

## 4.3 State Space Computation

A third optimization performs an image calculation as preprocessing step before unfolding the variable values to states in the transition from I/O-EFA to I/O-TS. The idea will be explained on a simple automaton, which has one input port u1 accepting integer numbers in the range [0; 249], one internal variable d1 having an integer range [2; 100], one output port y1, only one state A, and two transitions:

$A \xrightarrow[\{y1=5u1+d1;d1=2d1+u1\}]{[u1<d1 \wedge d1 \leq 5]} A$, $A \xrightarrow[\{y1=3u1\}]{[u1 \geq d1 \vee d1 > 5]} A$.

The second transition does not update any internal variables and does thus not affect the reachable state space in the I/O-TS. The data update function of the first transition is $f(d1) = 2 \cdot d1 + u1$ and the guard condition limits the range of d1 to [2,5] and u1 to [0; d1]. The algorithm computes a mapping M(f) of how internal variables' old values will be updated to new values. For the given I/O-EFA the mapping is $M(f) = \{2 \rightarrow \{4; 5\}; 3 \rightarrow \{6; 7; 8\}; 4 \rightarrow \{8; 9; 10; 11\}; 5 \rightarrow \{10; 11; 12; 13; 14\}\}$.

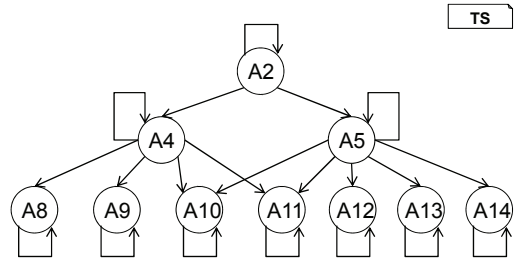

**Figure 10: Unfolded IO-TS (guard and output expressions are omitted for clarity).**

For the initial variable assignment d1 = 2 the unfolded I/O-TS as shown in Fig. 10 is created. The computation of image values of the data update functions in this example, has a great impact on the statespace of the I/O-TS. Unfolding the entire range of d1 ∈ [2; 100] would create 99 states while only ten of them (A2, A4, A5, A8 to A14) are reachable. In the current prototype, the mapping M(f) is calculated by "brute force exploration" on GPU chips. A static guard condition analyzer as preprocessing step minimizes the ranges for the GPU evaluation.

## 4.4 Extension: Fix Values on Free Ports

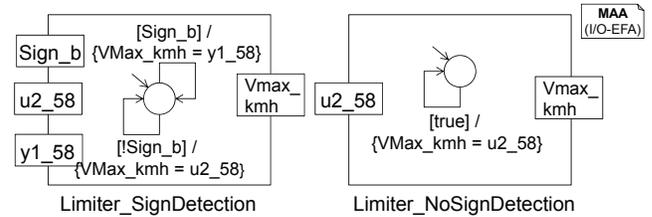

**Figure 11: Reduced IO-EFAs of Limiter Function variants in Fig. 8.**

The optimization from Subsect. 4.1 removes all syntactically identical elements on the control flow graph. Therefore only I/O-EFA from syntactically different CFGs are generated for comparison. Fig. 11 displays the reduced I/O-EFAs of the LimiterFunction component: The difference between both variants is that in the left one the feature *sign detection* is enabled and in the right one not.

This subsection introduces an extension that enables the comparison of Simulink models with different interfaces. The simulation algorithm in Subsect. 3.3 only works for similar component interfaces. To check backward compatibility of a component with an extended input interface, additional inputs might be required.

There are three general possibilities for providing inputs of extra ports: 1) There exists an adapter component calculating the values for the extra input ports based on the values of all available input ports. 2) Constant value blocks will be connected to the extra input ports. 3) Arbitrary (not constant) values can be connected to the extra input ports, because the extra ports have no behavioral impact on the reduced IO-EFA.

In initial experiments, the first option lead to complex adapter components resulting in a set of non-linear equations with many undefined functions, where often no solution could be found by the Z3-solver. The third option has the advantage that the extra input ports can be left blank and a direct replacement of the components is possible; but this leads to the large disadvantage in finding no behavioral compatible components, since typically all input ports have impact on the component's behavior. Thus the concept in this paper uses the second option as a trade-off.

In the example shown in Fig. 8, after assigning e.g. `false` to `Sign_b` and `zero` to `y1_58`, the behavior of the components with and without *sign detection* can be compared. Since the two automata are equal, the identity relation is a possible simulation preorder one. Therefore the sign detection component is behaviorally backwards compatible to the one without *sign detection* by fixing `Sign_b` to `false` and `y1_58` to `zero`. A more involved example will be used in the rest of this subsection to illustrate the extension.

A necessary condition of the simulation preorder relation existence is used to find the constant values for extra input ports: If *all unreachable states* of both automata *have been removed* and `B` is simulated by `A`, then every state of `B` must be in relation with at least one state of `A`. This result follows directly from the definition in Subsect. 2.2: If state $b$ is simulated by $a$ and a transition $e_b$ goes from $b$ to $b'$, then there must be a state $a'$ and a transition $e_a$ (being guard-compatible to $e_b$ as well as) going from $a$ to $a'$ such that $b'$ is simulated by $a'$. State $b$ is in relation with state $a$, if both states have the same output function according to the ranges of component `B`.

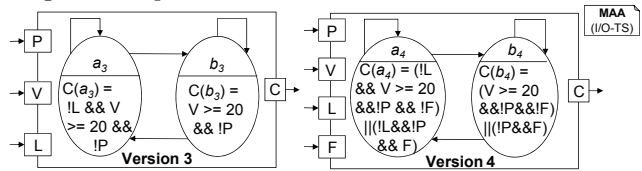

Figure 12: Shortened I/O-TS of `CruiseControl_On_Off` subcomponent (left: version 3, right: version 4). Transitions' guard labels are omitted.

Fig. 12 shows the strongly shortened I/O-TS of version 3 and 4 of the `CruiseControl_On_Off` subcomponent in the SPES demonstrator. Both versions have the input ports `ParkingBrake_Active (P)`, `Vehicle_Speed (V)`, `Limiter_Active (L)` and the output port `CruiseControl_Active (C)`. Version 4 has the extra input port `FollowToStop_Active (F)`. The necessary simulation preorder condition "every state of Version 3 must be in relation with at least one state of Version 4" requires both state $a_3$ and state $b_3$ being in relation with either $a_4$ or $b_4$. Eqn. (1) represents this condition for the following input port ranges: P, L, F are Boolean and V is an integer between 0 and 250 km/h.

$$\exists F \in B : \forall P, L \in B, V \in [0; 250] :$$
$$[C(A3) = C(A4) \vee C(A3) = C(B4)] \wedge$$
$$[C(B3) = C(A4) \vee C(B3) = C(B4)] \quad (1)$$

Using Z3's (`get-model`) command applied on the SMT-Expression in Eqn. (1) returns the constant value assignment `false` for F.

If the Z3 solver's model is empty, then there exist no constant values for the extra ports so that the necessary condition is satisfied. Thus component `A` is not behavioral backward compatible to component `B`.

Since there can be multiple vectors satisfying the necessary but violating the sufficient simulation preorder condition, the following steps may be repeated: fix the value on free ports (excluding already tested vectors) and check for a simulation relation.

This extension allows to do behavioral compatibility checks of Simulink models where one component interface is a subset of the other's one.

## 5. EVALUATION

The evaluation of the proposed concept has been performed on a simplified driver assistance system. This system is provided by Daimler AG in the context of research project SPES XT to support evaluation in relation to variability. Therefore the system focus on representing variability modeling in Simulink while simplifying the concrete functional behavior. In addition not only different variants of the system but also four different versions are provided by Daimler AG. As the prototype could only support a restricted set of Simulink blocks, before the prototype implementation we selected two specific subsystems to be used in the following case study.

These subsystems fulfill the following attributes: (1) Different variants or versions of the subsystem exist. (2) The different variants or versions provide positive and negative compatibility scenarios. (3) Different versions or variants provide a different level of syntactical equality. (4) The subsystem contains internal states. (5) The subsystem contains all elementary block types: `Switch`, `If-else`, `UnitDelay`, `ReadMemory`, `WriteMemory` as well as Boolean and arithmetic functions.

One of the selected subsystem is the `Limiter` Simulink block already illustrated in Fig. 8. It consists of 66 blocks. The second subsystem is part of the `CruiseControl` system (`CC`) and calculates the `CC` status (active / inactive). While the first subsystem is used to analyze the compatibility of different variants (Limiter with or without traffic sign detection), the second subsystem is used to compare different versions. The newest version (V4) consists of 33 blocks, while the oldest one (V1) consists of 26 blocks.

### 5.1 Results

In the following the results for each step are described in detail and finally a benchmark for the whole process is summarized.

In the first step the control flow extraction is performed. As can be seen in Tab. 1 the control flow extraction is per-formed quite fast from 132ms to 314ms and provides already a first rough overview on the subsystems complexity: the Limiter subsystem has nine and the CC has three internal variables. All three variables of CC are of type Boolean. In contrast the Limiter subsystem has also floating point variables, which are accessed via Write and Read blocks. Since it is possible to interpret these variables as global ones being modified by other subsystems before accessed by the Limiter, there is no need for internal variables of the analyzed subsystem anymore. This reduces the state space; the impact of these interpretation will be clear in latter evaluation parts.

Tab. 1 states for each system the related amount of nodes, edges and internal variables in the CFG. The fourth column shows the CFG extraction time in ms. Version 3 of CC and the different variants of Limiter are of similar size and their values are omitted in Tab. 1.

|  | nodes | edges | variables | time[†] |
|---|---|---|---|---|
| Limiter | 67 | 74 | 9 | 314 |
| CC (V4) | 34 | 38 | 3 | 142 |
| CC (V1) | 27 | 31 | 3 | 132 |

Table 1: Results of CFG extraction.

In the next step the clone detection is performed and finally the I/O-EFA is extracted. The results are shown in Tab. 2

|  | var. | trans. | subst.[†] | I/O-EFA[†] | sum[†] |
|---|---|---|---|---|---|
| Sum | - | - | 456 | 6403 | 6859 |
| Vmax | 5 | 15 | 456 | 4744 | 5200 |
|  | (6)[‡] | (50)[‡] |  | (74905)[‡] | (75361)[‡] |
| Active | 5 | 2 | 456 | 451 | 907 |
| CCSet | 2 | 2 | 456 | 130 | 586 |
| LimSet | 3 | 7 | 456 | 1078 | 1534 |
|  | (4)[‡] |  |  | (6787)[‡] | (6787)[‡] |
| Sum (r.) | - | - | 762 | 1193 | 1955 |
| Vmax (r.) | 0 | 2 | 762 | 1193 | 1955 |
| Active (r.) | 0 | 0 | 762 | 0 | 762 |
| CCSet (r.) | 0 | 0 | 762 | 0 | 762 |
| LimSet (r.) | 0 | 0 | 762 | 0 | 762 |
| CC(V4) | 3 | 3 | 135 | 1239 | 1374 |
| CC(V1) | 3 | 3 | 81 | 3588 | 3669 |
| CC(r.34) | 1 | 3 | 166 | 1839 | 2005 |
| CC(r.13) | 2 | 3 | 123 | 6787 | 6910 |

Table 2: Clone detection and I/O-EFA transformation results. From left: amount of variables and transitions of the resulting I/O EFA, duration for substitution and I/O-EFA transformation.

In the I/O-EFA transformation one automaton is created for each output port of the subsystem. Thus four automata were extracted from the Limiter subsystem. The first group of rows of Tab. 2 shows the results for the I/O-EFA transformation of the Limiter block without any syntactical clone detection, while the second group represents the results for a transformation after a clone detection has been performed (r. stands for reduced). For both subsystems a reduction of variables and transitions could be performed during the clone detection step, which also significantly reduces the I/O-EFA transformation step. In addition in the case of the Limiter subsystem it could already be identified that only one output port (Vmax_kmh) is influenced by a structural difference. In addition all internal variables could be removed. Comparing the interpretation of one floating point variable as global variable or as internal variable (in brackets), the transformation complexity already increased significantly.

---

[†]time in ms

[‡]Clamped values show internal variables not being considered as global ones.

The substitution is considerably faster than the following transformation and reductions implied during clone detection already significantly reduces the general duration.

|  | state | trans. | img.[†] | TS[†] | sum[†] |
|---|---|---|---|---|---|
| sum | - | - | - | - | 60434 |
| Vmax | 8 | 72 | 1497 | 55886 | 57383 |
| Active | 2 | 6 | 110 | 1330 | 1440 |
| CCSet | 3 | 3 | 2 | 112 | 114 |
| LimSet | 8 | 40 | 334 | 1163 | 1497 |
|  |  |  | (776962)[‡] | (-)[‡] |  |
| CC(V4) | 8 | 74 | 513 | 3595 | 4108 |
| CC(V1) | 8 | 56 | 1138 | 3356 | 4494 |
| CC(r.34) | 2 | 6 | 114 | 258 | 372 |
| CC(r.13) | 4 | 16 | 717 | 1487 | 2202 |

Table 3: Results transformation to transition system. From left: amount states and transitions of the resulting transition system, duration of image calculation and TS transformation.

In a last step the I/O-EFA is transformed to a transition system to remove all internal variables. The results are shown in Tab. 3. As the I/O-EFA of the Limiter subsystem does not contain any internal variables after the clone detection, only the results for the scenario, where no clone detection is used is shown.

As can be seen even if the resulting transition systems are quite small (8 states, 74 transitions) calculating the update function image and performing the transformation is still time intensive (up to a minute). The results in brackets demonstrate that a floating point internal variable could not be handled in a reasonable time slot. In addition it is shown that in each case the reduction via clone detection always reduces the overall transformation process.

| port | time[†] | result |
|---|---|---|
| Sum | 41016 | T[Sign_b = false]/F |
| Vmax | 24624 | T[Sign_b = false]/F |
| Active_b | 1761 | T/T |
| CCSet | 1051 | T/T |
| LimSet | 13580 | T/T |
| Sum(r.) | 644 | T[Sign_b = false]/F |
| Vmax(r.) | 644 | T[Sign_b = false]/F |
| Active(r.) | 0 | T/T |
| CCSet(r.) | 0 | T/T |
| LimSet(r.) | 0 | T/T |
| CC(34) | 30113 | T[FTS_active_b = false]/F |
| CC(13) | 14518 | F/F |
| CC(r34) | 2869 | T[FTS_active_b = false]/F |
| CC(r13) | 8341 | F/F |

Table 4: Results of simulation checks in both directions.

Finally in Tab. 4 the results of the simulation and the overall duration are shown. The boolean expression in brackets describe the condition under which a compatibility could be achieved. The variant of the component `Limiter` with the sign detection feature is fully compatible to the variant of the `Limiter` component regarding all outputs but `Vmax`. The calculation regarding `Vmax` is only upward compatible under the condition the variable `Sign_b` is false, which deactivates the sign detection temporarily. As result the variant with sign detection can safely replace each variant without the sign detection feature, if `Sign_b` is forced to false always. Similar results are provided for component `CC`. Between version one and version three of component `CC` exists no compatibility, while version four can replace version three safely as long as the variable `FTS_active_b` is set to false.

## 5.2 Discussion

The prototype only evaluated Simulink models with `Integer`, `Enumeration` or `Boolean` data types. The same concept is also possible for Simulink `fixed-point` [24] data types by storing the pre- and post-decimal digits of the fixed point number as two separate integers; a simple example is available at http://rise4fun.com/Z3/R5Ap. The problem using `fixed-point` data types is the dramatically increase of the state space introduced by these variable types.

As shown in the previous part of this section one main issue is unfolding internal variables, which is directly related to the state explosion problem. Extending the clone detection mechanism in the CFG by identifying clones of type *DF3* [13] or even semantic ones based on graph transformations as in [2] will possibly harness the state space explosion.

Another bottleneck of the algorithm is the image calculation. The usage of an intelligent evaluation order such as leftmost outermost together with substitution and graph reduction [34] will most-likely improve the image-calculation of the data update function on the GPU.

Another shortage of the presented concept are the many Z3 calls (which need on average about 20ms per call) to check whether the guard conditions of different transitions can imply each other in the simulation algorithm. Therefore a classification of the complex guard conditions based on the rank idea [10] of fast bisimulation would significantly reduce the number of guards to be compared.

Nevertheless both improvements will not remove the problem size regarding the state space explosion.

[21] explains how Simulink Stateflow charts can be transformed and integrated into the Simulink CFG in I/O-EFA syntax. The implementation of this transformation step is on progress right now. After it is finished variants and versions of Simulink components with Stateflow charts can be tested for behavioral compatibility.

## 6. RELATED WORK

There coexists a lot of work related to our proposal in terms of definitions of component compatibility, methods to formalize and compare behavior, works that analyze Simulink components, and approaches in the context of product line analysis.

Many types of component compatibility have been defined for various purposes, e.g., syntactic interface compatibility [5, 6], behavioral interface compatibility [11, 19], and various behavior equalities and refinements [6, 20, 30]. This paper employs an extension of syntactic interface compatibility that allows a special case of forward simulation [6], i.e., a modified syntactic interface with a mapping between ports and values. Replacing components requires full behavior compatibility. Many formalisms have notions of refinement similar to compatibility [6, 20, 30] that handle underspecification and uncertainty. For Simulink, component behavior is deterministic and the simulation relation is equivalent to trace containment [4].

Various target formalisms have been used to verify Simulink component behavior regarding to requirement specifications. Approaches include translations to (symbolic) model checkers, colored Petri nets, and I/O-EFA automata [26, 3, 17, 37]. For the purpose of this work a behavior comparison of the formal representation of two Simulink models is crucial – leading to the choice of I/O-EFA, a translation inspired by [37], and a simulation relation computation.

We are not aware of related work or any formalization of Simulink for the comparison of component behavior as presented in this paper.

A related line of work is semantic model differencing [22, 23, 18]. In semantic model differencing of behavior models, the result of a comparison is a set of witnesses for differences, e.g., an execution trace possible in one but not the other [23, 18]. The presented prototype for Simulink models computes one such trace in case of incompatibility. So far semantic differencing had not been applied to Simulink models.

The motivation for the comparison of Simulink models comes from the maintenance and evolution of software product lines. Many works analyze properties of products in product lines [35]. The presented approach is fundamentally different in that it compares behavior of components that may replace existing ones across products. Other works on product lines maintenance propose general refactorings [14] based on delta operations.

## 7. CONCLUSION

This paper presented a complete concept how behavioral compatibility checks of Simulink models can be accomplished. Thus the behavior of different component variants or versions can be full automatically compared.

In the introduced concept a lot of effort has been done to avoid the state explosion. Therefore an approach finding syntactical clones of type *DF2* [13] based on the CFGs in both models as well as unfolding I/O-EFAs to I/O-TSs based on the GPU's image calculations of the data update functions for internal variables has been described.

The evaluation part showed that the elimination of all possible internal variables (especially floating point ones) as soon as possible is necessary to keep the generated I/O-TSs small enough for further model checking analysis. The comparison of the two models can be performed in a significantly lower time period, if larger parts of the extracted CFGs are syntactical clones. In the context of software product line maintenance and evolution compatibilities between different variants and versions of components, thus similar models, needs to be considered. As consequence the proposed approach provides best performance in this context. The evaluation also revealed that it is possible to compare Simulink components with a common base and containing about hundred blocks in a few minutes on a commercial grade computer.

# 8. REFERENCES

[1] L. Aceto, A. Ingolfsdottir, and J. Srba. The Algorithmics of Bisimilarity. In *Advanced Topics in Bisimulation and Coinduction*. Cambridge University Press, 2011.

[2] B. Al-Batran, B. Schätz, and B. Hummel. Semantic Clone Detection for Model-based Development of Embedded Systems. In *MODELS*, pages 258–272. Springer, 2011.

[3] R. Alur, A. Kanade, S. Ramesh, and K. C. Shashidhar. Symbolic analysis for improving simulation coverage of Simulink/Stateflow models. In *EMSOFT*, 2008.

[4] C. Baier and J.-P. Katoen. *Principles of model checking*. MIT Press, 2008.

[5] M. Broy. (Inter-)Action Refinement: The Easy Way. In *Program Design Calculi*, volume 118 of *Series F: Computer and System Sciences*. Springer, 1993.

[6] M. Broy and K. Stølen. *Specification and Development of Interactive Systems. Focus on Streams, Interfaces and Refinement*. Springer, 2001.

[7] C. Cadar and K. Sen. Symbolic Execution for Software Testing: Three Decades Later. *Communications of the ACM*, 56(2):82–90, 2013.

[8] D. Cok, D. Deharbe, and T. Weber. Competition results for the UFNIA division. http://smtcomp.sourceforge.net/2014/results-UFNIA.shtml. Website accessed: 2014-03-19.

[9] D. Cok, A. Griggio, R. Bruttomesso, and M. Deters. The 2012 SMT Competition. In *SMT 2012*, volume 20 of *EPiC Series*. EasyChair, 2013.

[10] A. Dovier, C. Piazza, and A. Policriti. An Efficient Algorithm for Computing Bisimulation Equivalence. *Theoretical Computer Science*, 311(1–3):221 – 256, 2004.

[11] H. Foster, S. Uchitel, J. Magee, and J. Kramer. Compatibility Verification for Web Service Choreography. In *ICWS*, 2004.

[12] R. v. Glabbeek. The Linear Time-Branching Time Spectrum I - The Semantics of Concrete, Sequential Processes. In *Handbook of Process Algebra, chapter 1*. Elsevier, 2001.

[13] N. Gold, J. Krinke, M. Harman, and D. Binkley. Issues in Clone Classification for Dataflow Languages. In *IWSC*. ACM, 2010.

[14] A. Haber, H. Rendel, B. Rumpe, and I. Schaefer. Evolving Delta-Oriented Software Product Line Architectures. In *17th Monterey Workshop*, 2012.

[15] A. Haber, J. O. Ringert, and B. Rumpe. MontiArc - Architectural Modeling of Interactive Distributed and Cyber-Physical Systems. Technical Report AIB-2012-03, RWTH Aachen University, 2012.

[16] K. Holger, R. Bernhard, and V. Steven. MontiCore: a Framework for Compositional Development of Domain Specific Languages. *Int. J. on Software Tools for Technology Transfer*, 12(5):353–372, 2010.

[17] V. Januzaj and S. Kugele. Model Analysis via a Translation Schema to Coloured Petri Nets. In *PNSE*, 2009.

[18] P. Langer, T. Mayerhofer, and G. Kappel. Semantic Model Differencing Utilizing Behavioral Semantics Specifications. In *MODELS*, 2014.

[19] K. Larsen, U. Nyman, and A. Wąsowski. Modal I/O Automata for Interface and Product Line Theories. In *Programming Languages and Systems*, volume 4421 of *LNCS*. Springer Berlin Heidelberg, 2007.

[20] K. G. Larsen, U. Nyman, and A. Wasowski. On Modal Refinement and Consistency. In *CONCUR*, 2007.

[21] M. Li and R. Kumar. Stateflow to Extended Finite Automata Translation. In *COMPSACW*, 2011.

[22] S. Maoz, J. O. Ringert, and B. Rumpe. A Manifesto for Semantic Model Differencing. In *MoDELS Workshops*, 2010.

[23] S. Maoz, J. O. Ringert, and B. Rumpe. ADDiff: Semantic Differencing for Activity Diagrams. In *SIGSOFT FSE*, 2011.

[24] Mathworks. Simulink User's Guide. Technical Report R2015a, MATLAB & SIMULINK, 2015.

[25] R. Mayr. Process Rewrite Systems. *Information and Computation*, 156(1–2):264 – 286, 2000.

[26] S. P. Miller, M. W. Whalen, and D. D. Cofer. Software model checking takes off. *Commun. ACM*, 53(2):58–64, 2010.

[27] L. Moura and N. Bjørner. Z3: An Efficient SMT Solver. In *TACAS*, volume 4963 of *LNCS*. Springer, 2008.

[28] C. S. Păsăreanu and W. Visser. A Survey of new Trends in Symbolic Execution for Software Testing and Analysis. *Int. J. on Software Tools for Technology Transfer*, 11(4):339–353, 2009.

[29] M. Peter. Systematische Rekonfiguration eingebetteter softwarebasierter Fahrzeugsysteme auf Grundlage formalisierbarer Kompatibilitätsdokumentation und merkmalbasierter Komponentenmodellierung. In *Software Engineering 2013*, volume P-213 of *LNI*, 2013.

[30] J. O. Ringert and B. Rumpe. A Little Synopsis on Streams, Stream Processing Functions, and State-Based Stream Processing. *Int. J. Software and Informatics*, 5(1-2):29–53, 2011.

[31] J. O. Ringert, B. Rumpe, and A. Wortmann. *Architecture and Behavior Modeling of Cyber-Physical Systems with MontiArcAutomaton*. Number 20 in Aachener Informatik-Berichte, Software Engineering. Shaker Verlag, 2014.

[32] B. Rumpe. *Formale Methodik des Entwurfs verteilter objektorientierter Systeme*. Herbert Utz Verlag Wissenschaft, 1996.

[33] B. Rumpe. *Modellierung mit UML*, volume 2nd Edition. Springer, 2011.

[34] P. Sestoft. Demonstrating Lambda Calculus Reduction. In *The Essence of Computation*, volume 2566 of *LNCS*. Springer, 2002.

[35] T. Thüm, S. Apel, C. Kästner, I. Schaefer, and G. Saake. A Classification and Survey of Analysis Strategies for Software Product Lines. *ACM Comput. Surv.*, 47(1):6, 2014.

[36] F. Tip. A Survey of Program Slicing Techniques. *J. of Programming Languages*, 3(3):121–189, 1995.

[37] C. Zhou and R. Kumar. Semantic Translation of Simulink Diagrams to Input/Output Extended Finite Automata. *Discrete Event Dynamic Systems*, 22(2):223–247, 2012.